\documentclass[a4paper,twocolumn,11pt,unpublished]{quantumarticle} 
\pdfoutput=1
\usepackage[utf8]{inputenc}
\usepackage[english]{babel}
\usepackage[T1]{fontenc}
\usepackage{amsmath}
\usepackage{graphics}
\usepackage{graphicx}
\usepackage{dcolumn}
\usepackage{hyperref}
\usepackage{tikz}
\usepackage{lipsum}
\usepackage{braket}\usepackage{graphicx}
\usepackage[english]{babel}
\usepackage{amsmath,amsfonts,amssymb,bbm}
\usepackage{enumerate}
\usepackage{textcomp}
\usepackage{listings}

\newcommand{\trace}{\mathrm{Tr}}
\newcommand{\vect}{\mathrm{vec}}

\newcommand{\QST}{\mathbb{QST}}
\newcommand{\IQP}{\mathbb{IQP}}

\newcommand{\sig}{\bs{\sigma}}
\newcommand{\Sig}{\bs{\Sigma}}
\newcommand*{\bs}[1]{\boldsymbol{#1}}
\newcommand{\norm}[1]{\left\lVert#1\right\rVert}

\DeclareMathOperator{\im}{\mathring \imath}
\DeclareMathOperator*{\softmax}{softmax}
\newcommand{\zero}{\ket{0}}
\newcommand{\one}{\ket{1}}

\newtheorem{definition}{Definition}
\newtheorem{assumption}{Assumption}

\usepackage{framed,color,verbatim}
\definecolor{shadecolor}{rgb}{.9, .9, .9}

   {\snugshade\verbatim}%
   {\endverbatim\endsnugshade}

\usepackage{mathtools}
\newcommand\SmallMatrix[1]{{%
  \small\arraycolsep=0.7\arraycolsep\ensuremath{\begin{pmatrix}#1\end{pmatrix}}}}

\begin{document}

\title{Quantum State Tomography as a Bilevel Problem}

\author{Georgios Korpas}
\email{georgios.korpas@fel.cvut.cz}
\author{Jakub Marecek}
\email{jakub.marecek@fel.cvut.cz}
\affiliation{%
Department of Computer Science, 
Czech Technical University in Prague, Karlovo nam. 13, Prague 2, Czech Republic}


\maketitle

\begin{abstract}
  It is natural to ask how to utilize {actual} measurements, such as the so-called IQ-plane data obtained in the dispersive readout of transmon qubits, in the estimation of the state of a quantum system. 
We formulate the joint problem of discrimination and quantum state tomography as a bilevel optimization problem
and show how to solve it. 
The use of the joint problem can improve the sample complexity (or the reconstruction error for a fixed number of measurements) 
compared with traditional techniques that decompose the problem into the discrimination and state tomography based on 
the estimated expectation values of certain projective measurement operators.
\end{abstract}

\section{Introduction and motivation}\label{sec1:level1}
\begin{figure}[b]
    \centering
    \includegraphics[scale=0.7]{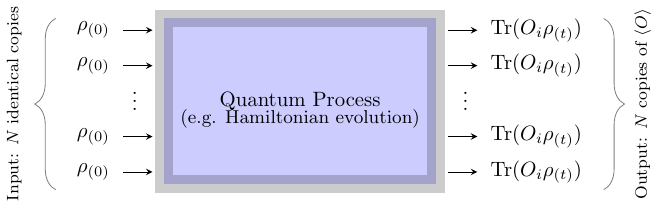}
    \caption{Several copies of a known input state $\rho_{(0)}$ are fed into a quantum process and in each iteration a measurement of some observable $O_i$ is performed. These outputs then are combined to form an empirical average of $\braket{O_i}$ out of which the output states can be reconstructed. The state reconstruction part is usually performed with heuristic techniques such as maximum likelihood estimation.}
    \label{fig:qst}
\end{figure}

The development and validation of small but nontrivial quantum devices have already had a significant impact, e.g. \cite{Montanaro_2016,Wendin_2017,Preskill_2018,Arute:2019zxq}. However, further progress is required to validate, benchmark, and fully exploit such quantum devices. 
In general, quantum system identification refers to the collections of techniques and protocols employed for this purpose \cite{NielsenChuang}. 
At the most basic, one wishes to reconstruct a state's density matrix from 
the measurement of an ensemble of copies of a quantum state, associated with a quantum device of interest, which is known \cite{NielsenChuang,MauroDAriano2003} as quantum state tomography (QST).
A QST schematic is presented in Fig. \ref{fig:qst}.

QST can be performed using many classical techniques, such as maximum likelihood estimation, compressed sensing, and Bayesian state tomography, among others. 
The development of such techniques is a subject of intense research, for example, in the context of matrix product states, \cite{Cramer2010}, using neural networks \cite{Quek2018}, optimizing circuits for noisy gates \cite{Ivanova2022}, while recently a great deal of attention has been paid to the use of the classical shadow \cite{Huang2020}. Although these techniques are well established, some issues remain. The key issue is that these techniques require heavily preprocessed data as input. To be more precise, the output measurements are often considered at a very abstract level: 
the empirical estimates of 
the expectation values of certain projective measurement operators.
These can be obtained only at the end of a nontrivial measurement chain involved in the readout of the quantum devices and with a substantial amount of signal processing. For example, superconducting qubits \cite{wendin2017quantum, PhysRevApplied.7.054020}, and increasingly also quantum dots/spin qubits \cite{schaal2020fast}, tend to utilize the so-called dispersive readout, which we explain in further detail in the following, but which produces a complex-valued signal in response to a so-called readout pulse \cite{McClure2016}. 
This signal is processed into the so-called IQ-plane data, which are complex valued.
The empirical estimates of the expectation values are then obtained from the IQ-plane data using the so-called discrimination procedure. 


In this paper, we show how to perform QST from raw data, prior to discrimination. 
Specifically, our main result is the problem formulation \eqref{SDPdensity2}, which, at a high level, reads
\begin{equation}\label{SDPdensity0}
\begin{aligned}
    \arg \min_{b, {\widehat {\bs \rho}}} &\quad \norm{\bs \sum_{I}{\rm Tr}(\widehat {\bs\rho}\bs \sigma_I) - b}_{2}^2 \\
    {\text{subject to}} & \quad b  \in \text{ set of optimizers of a lower} \\
                        & \quad \quad \quad  \text{  level optimization problem}\\
                        & \quad \trace{(\widehat {\bs\rho})} = 1 \\
                        & \quad \widehat{\bs \rho} \in \mathbb{S}_+^2.                    
\end{aligned}
\end{equation}
where $\bs \sigma_I \in \{ \bs \sigma_x,\bs \sigma_y,\bs \sigma_z \}$.
That is, in \eqref{SDPdensity0} we minimize the empirical risk, where $b$ is one of the optimal solutions to a lower-level minimization problem associated with the 
discrimination between basis states using the raw data.  
This can be seen as a joint problem of estimating the density matrix from the measurements and estimating the measurements
from the raw data. In mathematical optimization, this is known as a bilevel optimization problem. 

In contrast, processing the IQ-plane data into empirical estimates of the expectation values of projective measurement operators in a discrimination procedure, followed by a QST procedure utilizing empirical estimates of the expectation values, can be seen as a decomposition of the joint problem.
Such a decomposition necessarily increases the overall sample complexity and may also introduce non-Gaussian artifacts in the data.
This is the case especially when the procedures in the decomposed approach are suboptimal, as is the case with commonly used heuristics, such as the expectation–maximization (EM) algorithm in the discrimination and least-squares approaches that ignore the semidefiniteness of the density matrix within the QST utilizing the empirical estimates of the expectation values of projective measurement operators. 

Note that the sample complexity of QST depends on the precise reconstruction method and the type of measurements performed. 
Although there are information-theoretic arguments showing \cite{Haah2016} that certain algorithms are optimal with respect to the number of samples needed for the last step of the decomposition, assuming that the discrimination process is performed without any errors.
In practice, there are errors that propagate from discrimination to quantum state tomography in a fashion that cannot be controlled in the traditional, decomposed approach. For example, when measuring the system state in superconducting devices, part of the erroneous results can be due to qubit crosstalk effects which propagate to the readout fidelity and special care must be taken so as to tackle this \cite{Duan2021}. 

The joint problem, which utilizes the raw data (such as the IQ-plane data or the dispersive readout signal directly, cf. Sec. \ref{sec:sec2}), is capable of reducing noise and its propagation, essentially by filtering out samples of noise that contaminated the raw data, with the objective of minimizing the empirical risk in QST. This seems hard to do in a principled fashion without considering the use of the results of the discrimination.
As we illustrate in Sec.~\ref{sec:example},
the joint problem makes it possible to obtain better estimates, given a certain number of samples, than using the decomposition approach. The lower sample complexity, in turn, translates to the computational efficiency of the reconstruction method; it is not uncommon to have weeks of postprocessing time for QST of non-trivial devices. 

Progress in QST may have a considerable impact on progress in quantum computing in a more general sense. 
QST is the current \emph{de facto} standard for the characterization and verification of quantum devices, 
including many randomized benchmarking procedures. 
For example, to implement quantum gates, one needs to characterize the operation of a quantum device by running a series of known inputs and reconstructing the corresponding outputs using QST, such as in the fidelity estimation of CNOT gates \cite{cnot} using QST. 
Thus, the role of QST in quantum technologies is of fundamental importance.

\section{Dispersive Readout of a Qubit}\label{sec:sec2}

QST requires the acquisition of data from the quantum device that is investigated, which must be well isolated from sources of noise or dissipation from their coupled environment. 
Popular quantum devices that satisfy the above criteria are superconducting qubits such as transmon qubits \cite{Koch_2007}, which have been popularized by IBM and Rigetti Computing, as well as xmon qubits used by Google \cite{GoogleReadout}. A common practice is to couple the qubit(s) to a dispersive oscillator that has a resonant frequency that depends on the qubit state. By probing it with a pulse \cite{McClure2016} and reading and analyzing the response pulse, one can reveal nontrivial information about the state.

The dispersive readout of the qubit refers to the process of determining whether the qubit was measured in the $\ket{0}$ or the $\ket{1}$ eigenstate with respect to the measurement operator $\bs M$. To determine the qubit state using the readout signal as the bare minimum information on the quantum state, a number of steps are taken in the readout chain.
The readout chain is made up of three levels of increasing complexity, which are commonly \cite{alexander2020qiskit} known as:
\begin{enumerate}\addtocounter{enumi}{-1}
    \item \emph{raw data} correspond to a discrete-time signal of the output pulse with frequency $\omega_{\rm r.o.}$.
    \item \emph{IQ-plane data} corresponds to removing the frequency component of the readout signal (see Eq. \ref{IQeq}) and obtaining the complex valued in-phase and quadrature component (IQ) data.
    \item \emph{discriminated data} which are obtained by applying a discrimination procedure to the IQ-plane data.
    From the discriminated data, we can obtain the so-called $b$-vector, cf. Eq. \ref{bi}.
\end{enumerate}
Only the output of the final step, i.e. the output of the discrimination procedure, is passed to QST routines. 

\begin{figure}[t]
    \centering
    \includegraphics[width=1.0\columnwidth]{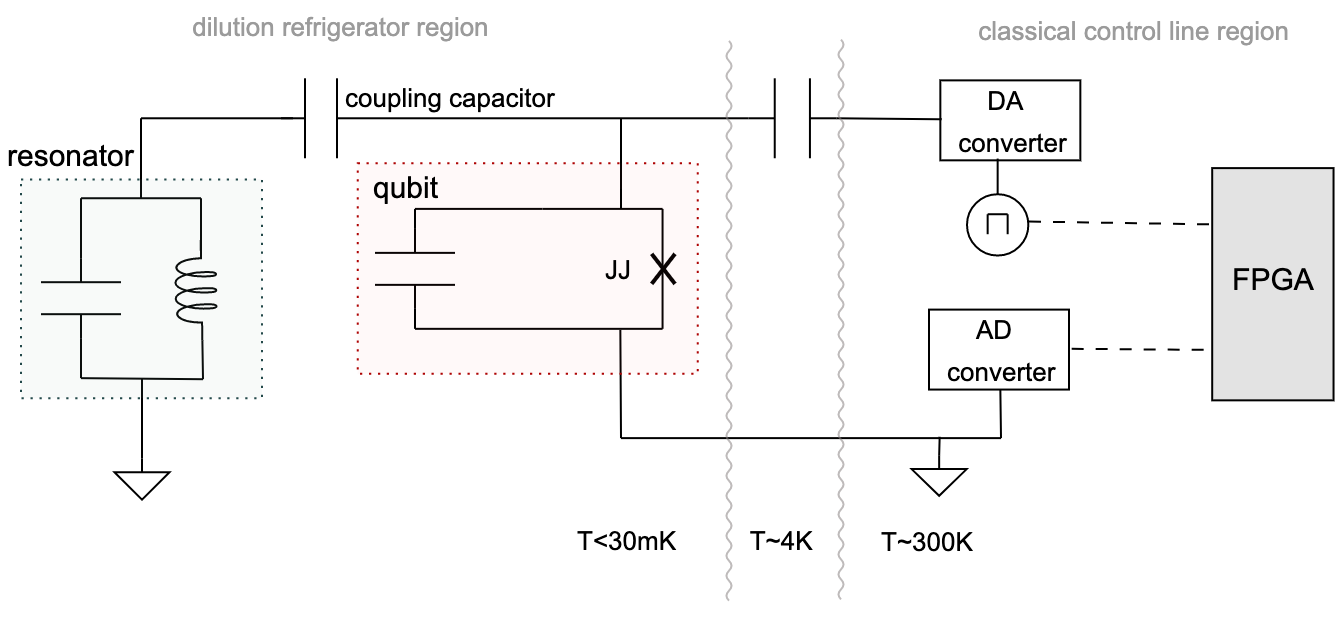}
    \caption{Simplified schematic of the transmon qubit device realized by a Josephson junction coupled to an LC resonator via a coupling capacitor. Similarly, the cavity which is kept at $T<10$mK (it can be lower than 3mK), is coupled to the control line via another weak coupling capacitor. The input pulse from the classical control line is converted to an analog signal, which interacts with the qubit. The output pulse passes through an AD converter to the FPGA controller interface where the I-Q readout is recorded. In this article we suggest perfroming QST precicely from this data, see Sec. \ref{sec:bilevel} and Eq. \eqref{SDPdensity2}.    
    }
    \label{qubit_cavity_control}
\end{figure}

\begin{figure*}[t]
    \centering
    \includegraphics[width=1\textwidth]{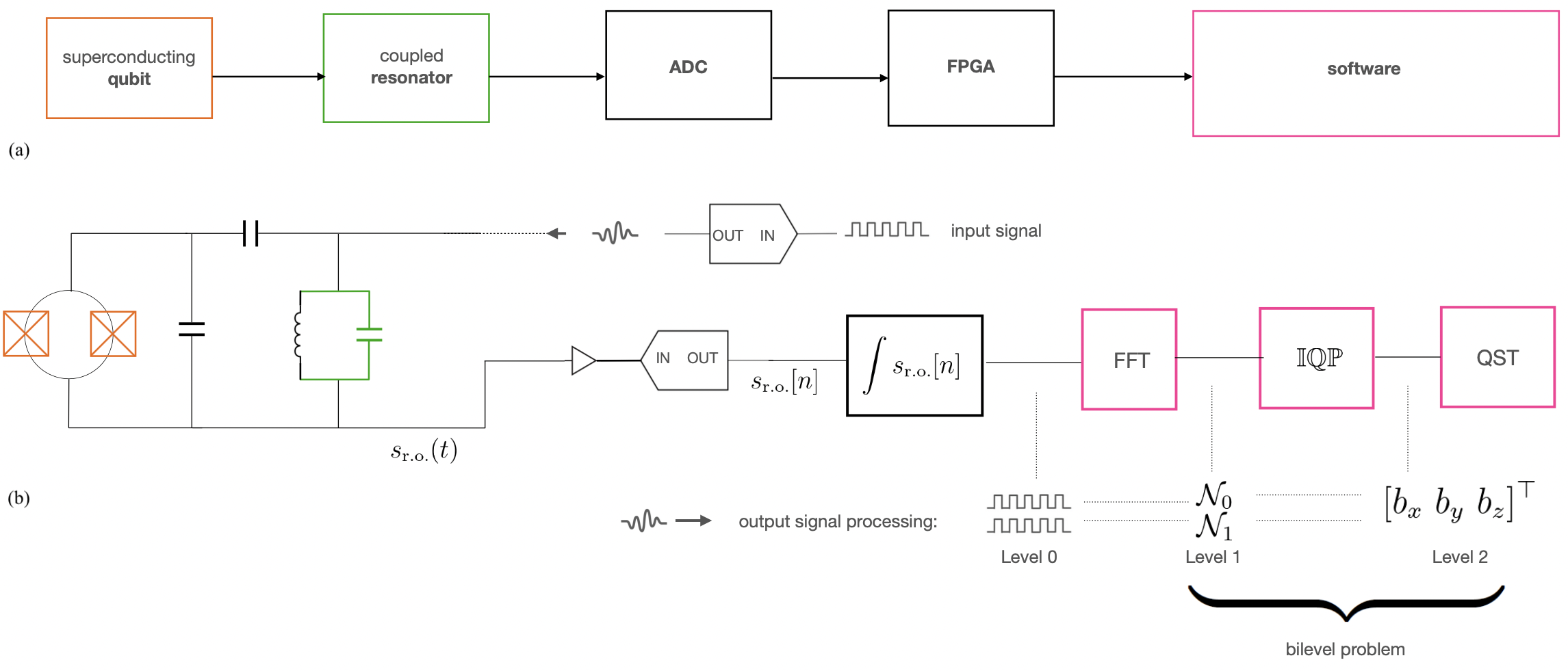}
    \caption{(a) A flow of operations in the signal readout. (b) A simplified schematic of the transmon qubit device realized by a Josephson junction coupled to an LC resonator via a coupling capacitor.
    See \cite{vandijk2019electronic} for a  detailed analysis of the structure. There are a number of steps in the output signal processing; our suggestion is to replace the discriminator ($\mathbb{IQP}$, Eq. \ref{GMM}) and the traditional quantum state tomography ($\QST$, \ref{SDPdensity}) with the bilevel problem \eqref{SDPdensity2}. See App. \ref{app:readout} for further details.}
    \label{qubit_cavity_control}
\end{figure*}

At the raw data level, by analyzing the integrated return pulse $\omega_{\rm r.o.}$, one can deduce whether the quantum device is in the ground state or in the excited state on the pointer basis $\{\ket{e},\ket{g}\}$. 
For the purposes of this article, we can safely approximate $\ket{g}\approx \ket{0}$ and $\ket{e} \approx \ket{1}$, which correspond to the eigenstates of the observable of the measurement apparatus, as discussed in \cite{Zurek:1981xq}. The raw data can be mapped to the phase space (at level 1), whose precise meaning is explained below, where the output response pulse (at level 0) is represented by the amplitude response $I$ and the phase response $Q$. 
Repetition of the measurement $n$ times produces a mixture of two distributions in the IQ plane. 

The IQ plane, which we utilize throughout this article, can be thought of as the phase space of the resonator-qubit coupled system. 
A comparison of the readout pulse with the original pulse using the phase shift as a result of the measurement allows one to map the qubit state onto the IQ plane.
In particular, after probing the transmon qubit with a linear signal \cite{McClure2016} of frequency $\omega_{\rm probe}$, one reads out the response pulse $s_{\rm r.o.}(t) = {\rm Re}(A_{\rm r.o.}\exp(i(\omega_{\rm r.o.}t + \vartheta)))$, where $(A_{\rm r.o.}, \omega_{\rm r.o.}, \vartheta)$ correspond to the amplitude, frequency, and phase of the readout pulse, respectively. 
At a fixed frequency $\omega_{\rm r.o.}$, the phasor can be expressed as $s|_{\omega_{\rm r.o.}} = {\rm Re}(A\exp{\im \vartheta})$, which completely specifies the pulse at the given frequency. Finally, we record the \emph{in-phase} component $I$ and the \emph{quadrature} component $Q$ of the phasor:
\begin{equation}\label{IQeq}
    \begin{aligned}
    s|_{\omega_{\rm r.o.}} &= {\rm Re}(A\exp{\im \vartheta}) \\
                           &= I + \im Q,
\end{aligned}
\end{equation}
which are known as the IQ-plane data.

 In total, $n$ repetitions of the measurement allow the computation of the empirical estimate of the expectation value of the measurement operator. On a $N$-dimensional measurement basis with $N$ measurement operators,
  the repetition of the same procedure with further copies of the measured state  
 (albeit possibly with different pulse frequencies) makes it possible to prepare a histogram approximation of a $N$-dimensional random variable, which is the input to traditional QST routines.

\section{QST as bilevel SDP}\label{sec:bilevel}

Let us consider an example of the QST on a single qubit. To perform QST, one prepares an ensemble of $n_I$ identical states $\bs \rho$ and performs the measurement of Pauli observables $\sig_I$ of each of these copies.
Breaking away from the readout chain of the previous section, 
we obtain $n_I^{\zero}$ measurements corresponding to the $\ket{0}$ state and $n_I^{\one}$ measurements corresponding to the $\ket{1}$ state where $n_I = n_I^{\zero} + n_I^{\one}$.
This corresponds to a binary vector $\boldsymbol{\beta}_I \in \{0,1\}^{n_I} $.
The three measurement observables of the Pauli basis provide us with three such vectors ($\boldsymbol{\beta}_x,\boldsymbol{\beta}_y, \boldsymbol{\beta}_z$). Out of each of these vectors, one can write the empirical estimates $b_I$ for the expectation value of $\sig_I$ as:
\begin{align}\label{bi}
    b_I = \frac{1}{n_I}\left (\sum_{j\in n_I^{\ket{0}}} (\boldsymbol{\beta}_I)_j -\sum_{j\in n_I^{\ket{1}}} (\boldsymbol{\beta}_I)_j \right).
\end{align}
For a single qubit and for $n = n_x + n_y + n_z$ measurements in the Pauli basis, we obtain the following empirical estimates for the expectation values of the Pauli observables:
\begin{align}\label{bvector}
    b = (b_x \,\,\, b_y \,\,\, b_z)^\top \in [-1,1]^3,
\end{align}
i.e., within a cube. 

Traditional QST procedures use the $b$ vector to estimate the density matrix.
Thus, QST can be interpreted as a function from the 
 polyhedron $[-1,1]^M$ to
 the space $\mathbb{H}_1^n$ of $n\times n$ Hermitian matrices with unity trace:
\begin{equation}\label{eq:QST0}
    \QST : [-1,1]^M \to \mathbb{H}_1^n.
\end{equation}
Therein, convex optimization methods 
provide \emph{shape-constrained least-squares} fit.
In particular, the objective of the constrained least-squares problem is to find a unitary Hermitian matrix $\widehat{\bs\rho} \in \mathbb{H}_1^2$, which estimates the density matrix of interest, such that the $\ell_2$-norm
\begin{align}\label{IBMopt}
 \norm{\bs A\vect(\widehat {\bs \rho}) - b}_{2}^2,
\end{align}
is minimized, where:
\begin{itemize}
    \item Matrix $\bs A$ stacks the vectorized measurement operators. For a qubit and measurement operators in the Pauli basis:
        \begin{align*}
            \bs A = \begin{bmatrix}
                    \vect(\sig_x) \,  \vect(\sig_y) \, \vect(\sig_z)
                \end{bmatrix}^\top  \in \mathbb{C}^{4\times 3},
        \end{align*}
    \item Vector $b$ corresponds to the vector of empirical estimates of Equation (\ref{bvector}).
\end{itemize}


Shape-constrained least squares (\ref{IBMopt}) can be 
formulated as a semidefinite programming (SDP) problem \cite{wolkowicz2012handbook}:

\begin{equation}\label{SDPdensity}
\begin{aligned}
    \QST_{\bs A}(b) := \arg \min_{\quad {\widehat{\bs\rho}}\, \succeq 0} &\quad \norm{\bs A \vect(\widehat{\bs\rho}) - b}_{2}^2 \\
    {\text{subject to}} & \quad \trace{(\widehat{\bs\rho})} = 1. \\
                        & \quad \widehat{\bs\rho} \in \mathbb{S}_+^2, \
\end{aligned}
\end{equation}
where $\mathbb{S}^n_+$ denotes the space of positive semidefinite matrices with complex values. 
Solving the SDP \eqref{SDPdensity} can be seen as a map from the space of recorded measurements $b$ to the space of estimates $\widehat{\bs \rho}$,
as suggested in Equation (\ref{eq:QST0}).
Problem (\ref{SDPdensity}) is easily generalized to (i) other measurement basis choices, as well as to (ii) higher-level systems with the corresponding generalization of $\bs A$ and $b$.


\subsection{Bilevel formulation of QST}
\label{sec:bilevelSDP}

In the preceding example and in Prob. (\ref{SDPdensity}), we assume that the vector $b$ is given.
Instead, we could start with the IQ-plane data and a map $\IQP_{\mathcal{D}_{\ket{0},\ket{1}}}(u)]_I$ from the IQ-plane data to the $b$-vector, which we formalize subsequently. 
Then, we can perform QST directly using the IQ-plane data by reformulating Prob. (\ref{SDPdensity}) as a bilevel problem:

\begin{equation}\label{SDPdensity2}
\begin{aligned}
    \QST_{\bs A}(b) := \arg \min_{b, {\widehat {\bs \rho}}\, \succeq 0} &\quad \norm{\bs A \vect(\widehat {\bs\rho}) - b}_{2}^2 \\
    {\text{subject to}} & \quad b_I \in [\IQP_{\mathcal{D}_{\ket{0},\ket{1}}}(u)]_I \\
                        & \quad \trace{(\widehat {\bs\rho})} = 1 \\
                        & \quad \widehat{\bs \rho} \in \mathbb{S}_+^2.                    
\end{aligned}
\end{equation}
where $u=\{x,y,z\}$. In the new constraint (\ref{SDPdensity2}), $b_I$ is defined to belong to the set of optimizers of the so-called lower-level optimization problem, which we introduce next. 

Returning to the IQ-plane data, let us fix a measurement observable $\bs M$ and let $\Omega_{\bs M}= \Omega_{\bs M}^{\zero}\cup \Omega_{\bs M}^{\one}$ denote the space of possible outcomes of the measurement of the quantum device with respect to $\bs M$. $\Omega_{\bs M}$ is indeed the union of the outcomes that will be labeled as belonging to the $\ket{0}$ cluster, and the ones that will be labeled as belonging to the $\ket{1}$ cluster. Assuming that the recorded samples belong to a mixture of two Gaussian probability distributions $\mathcal{D}_{\ket{0}} \sim \mathcal{N}(\mu_0,\Sig_0)$ and $\mathcal{D}_{\ket{1}}\sim \mathcal{N}(\mu_1,\Sig_1)$, where $\mu_j$ is the mean of the $j$-th distribution and $\Sig_j$ is the covariance matrix of the $j$-th distribution, the $b$ vector can be interpreted as the evaluation of a map ``\text{IQ-plane data} $\mapsto b_I$'':
\begin{align}\label{eq:IQP}
    \IQP_I:& (\Omega_{\bs \sigma_I}; \mathcal{D}_{\ket{0}}, \mathcal{D}_{\ket{1}}) \to [-1,1], 
\end{align}
the evaluation of which produces the $I$-th component of the $b$ vector, the component corresponding to $\bs \sigma_I$. Here, we index $\IQP$ to stress the fact that for the full QST, one needs to perform this operation for all possible measurement operators of the measurement basis.

In the Huber contamination model, there are two distributions with parameters $\theta_0 = \{\alpha_0, \mu_0, {\bs \Sigma}_0\}$ corresponding to the $\ket{0}$ state, and parameters $\theta_1 = \{\alpha_1, \mu_1, {\bs\Sigma}_1\}$ corresponding to the $\ket{1}$ state, and the mixing coefficient $\alpha_3$ for an unknown arbitrary distribution $g(x)$ corresponding to adversary noise. 
Then, $\IQP$ of Equation (\ref{eq:IQP}) corresponds to 

\begin{align}
\label{dis_adv}
 f(u_s) & = \sum_{i=\{1,2\}}   \mathcal{D}_{\ket{i}}(u_s)  +\alpha_2 g(u_s) \\
\textrm{where } \mathcal{D}_{\ket{i}}(u_s) & :=  \frac{1}{\sqrt{2\pi}} \frac{\alpha_i}{|\Sig_i|^\frac{1}{2}} e^{(u_s - {\mu}_i)^\top \Sig_i^{-1}(u_s-{\mu}_i)}, \notag
\end{align}
with $\alpha_0+\alpha_1+\alpha_2=1$, $\alpha_i \in [0,1], \, \forall i \in \{0,1,2\}$, and $u_s$ is the random variable corresponding to the sample $s$. We can now define $\IQP$ from Equation (\ref{eq:IQP}) exactly as follows. First, we define:
\begin{equation*}
  f_{\IQP} := \sum_{s=1}^{|S_I|} \left[\sum_{i \in\{0,1 \}} c_s^{\ket{i}} \log \mathcal{D}_{\ket{i}}(u_s) + c_s^{\varepsilon}g(u_s)  \right]
\end{equation*}
Let $|S_I| \equiv n_{I}$ denote the number of recorded samples for the $i$-th measurement operator. Then, for the feasible set $U = \{ c_s^{\ket{0}}, c_s^{\ket{1}},c_s^{\rm noise} \in \{ 0, 1 \},  \mu_1,  \mu_2,\Sig_1, \Sig_2, \alpha_1, \alpha_2$\}, $\IQP$ corresponds to the following nonconvex problem:
\begin{equation}
    \begin{aligned}\label{GMM}
    \IQP_{\mathcal{D}_{{\zero},{\one}}}(u_s) :=\arg \min_{U} & \quad
    f_{\IQP}
    \\ \noindent 
    \textrm{ s.t. } &  \sum_{s=1}^{|S_I|} c_s^{\ket{i}} = \alpha_i |S_I| \\ 
    & 1 = \sum_{j=0}^2\alpha_{i}.
\end{aligned}
\end{equation}

Notice that there is no need to perform the inversion for $\Sig_i^{-1}$: one can optimize the matrix variable, whose meaning is to be the inverse of the covariance.

Refining the interpretation of \eqref{GMM} as a map from the IQ-plane data corresponding to the $\bs \sigma_I$ measurement observable to the space of $\boldsymbol{b}_I$ we have:
\begin{equation}\label{boldb}
\begin{aligned}
    \IQP_I &:  (\Omega_{\bs \sigma_I}; \mathcal{D}_{\ket{0}}, \mathcal{D}_{\ket{1}})  \to [-1,1], \\
         &  u_s \mapsto (\boldsymbol{b}_I)_s. 
\end{aligned}
\end{equation}
Both the bilevel problem (\ref{SDPdensity2}) and its lower level problem (\ref{GMM}) are not trivial. However, in some settings, significant progress can be made. 

First, let us consider the simplest setting, where the parameters of Equation (\ref{dis_adv}) are known 
and either there is no contamination ($\alpha_2 = 0$) or the contaminated samples can be identified. When one considers this ``noise-less'' case, Problem (\ref{GMM}) corresponds to the well-known problem of the degree of membership (posterior information) of the samples, whose simplest case is:
\begin{equation}
 \begin{aligned} \label{beta_arg_max}
\bs \beta_{\ket{i}}(u_s) & = \arg \max_{i\in \{0,1\}} N \Big\{\bs X_i(u_s) \Big\}, \
\end{aligned}  
\end{equation}
with $\bs \beta_{\ket{i}}$ relating to the $b$ vector as prescribed in Equation \eqref{bi}, although the role of the index is different in this context: we fix $\bs \beta_{\ket{i}} :=  (\bs \beta_I)_{\ket{i}}$, for some $I$. Furthermore, $N$ is a normalization factor that we set equal to one, and
\begin{align}\label{X}
    \bs X_i(u_s) := -(u_s-\mu_i)^\top { \Sig}_i^{-1}(u_s-\mu_i),
\end{align}
and the decision rule is to assign each sample $u_s$ to the cluster whose mean minimizes the Mahalanobis distance. Prob. \eqref{beta_arg_max} has a solution that is computable in linear time. Note that by expanding the terms in \eqref{X}, we can define a PSD matrix
\begin{align}
    \bs F^{(u)}_{\ket{i}} = \begin{pmatrix}
           \mu_i^\top\Sig_i^{-1} \; - \; \mu_i^{\top}\Sig_i^{-1} \mu_i \
        \end{pmatrix}.
\end{align}
Then \eqref{beta_arg_max} can be written as:
\begin{align}
    \bs \beta_{\ket{i}}(u_s) &= \arg \max_{i\in\{0,1\}} \,\,\,\begin{pmatrix} \tilde u_s^\top &1 \end{pmatrix} \bs F^{(u)}_{\ket{i}} \tilde u_s,
\end{align}
where $\tilde u_s :=  (u_s^\top \otimes_{\rm Kron}1 )^\top$. Alternatively:
\begin{align}
    \bs \gamma_{\ket{i}}(u_s) &=  \softmax_i \,\,\,\begin{pmatrix} u_s^\top &1 \end{pmatrix} \bs F^{(x)}_{\ket{i}} \begin{pmatrix} u_s\\1 \end{pmatrix},
\end{align}
where $\softmax$ is defined as
\begin{align}
    \softmax_i(u_s) = \frac{\exp(y_i(u_s))}{\exp( y_0(u_s) ) + \exp(y_1(u_s))},
\end{align}
such that $\bs \gamma_{\ket{0}}^{(u_s)}+ \bs \gamma_{\ket{1}}^{(u_s)}= 1$,
where, by definition, $\bs \gamma_{\ket{i}}(u_s)$ will be in the interval $[0,1]$.

When $\bs F_{\ket{i}}$ are nonnegenerate, this yields a strongly convex, unconstrained 
lower-level problem, which in turn can be substituted by its first-order optimality 
conditions. 
Then, it is possible to formulate the SDP \eqref{SDPdensity2} with additional constraints, taking into account all measurement basis operators.
Explicitly, we have:

\begin{align}
    \QST_{\bs A} := \arg \min_{\quad {b, \widehat{\bs\rho}}, \bs \gamma} &\quad \norm{\bs A \vect(\widehat{\bs\rho}) - b}_{2}^2 \label{SDPdensity3} \\
    {\text{subject to}} & \quad \trace{(\widehat{\bs\rho})} = 1 \notag  \\
                        & \quad \widehat{\bs\rho} \in \mathbb{S}_+^2 \notag  \\
                        & \quad b = (b_x \,\, b_y \,\, b_z)^\top \notag \\
                        & \quad b_u = \gamma_{\ket{0}}^{(u_s)} - \gamma_{\ket{1}}^{(u_s)} \notag  \\
                        & \quad   \gamma_{\ket{i}}^{(u)} = \sum_{u_s}\softmax_i \tilde u_s^\top \bs F^{(u)}_{\ket{i}}\tilde u_s \notag 
\end{align}
%
for all elements $u_s$. This bilevel SDP \eqref{SDPdensity3} can then be extended, for instance, toward the unknown parameters of the Gaussian mixture and 
adversarial noise.

When the mixture model is contaminated by noise, that is, $\alpha_2 > 0$, we need to consider a constrained optimization problem. 
When the mean vectors $\mu_i$ of Equation (\ref{dis_adv}) are unknown but the covariance matrices ${\Sig}_i$ are known, the estimation of the mean vectors $\hat{\mu}_i$ and mixture weights $\hat{a}_i$ in the lower-level problem (\ref{GMM}) corresponds to a nonconvex, but commutative polynomial optimization problem. Although this makes the problem nontrivial, it has been studied \cite{jeyakumar2016convergent}. 
Finally, when both the mean vector and the covariance matrices of Equation (\ref{dis_adv}) are unknown and there is contamination ($\alpha_2 > 0$), the lower-level problem (\ref{GMM}) then corresponds to a nonconvex noncommutative (i.e., operator-valued) polynomial optimization problem. One can again solve such a bilevel polynomial optimization problem with a nonconvex lower-level problem using hierarchies of semidefinite programming relaxations. 
All of these 
more complicated scenarios are discussed in the supplementary material.

\section{A simple example}
\label{sec:example}

To illustrate the importance of solving the joint problem, rather than decomposing it into discrimination and quantum state estimation, let us consider an example in which a state is estimated using information from input-output data in the presence of normally distributed noise with unity covariance and mean $\mu_{\rm noise} = (-3.5,-3.5)$. Table \ref{T1} shows that the estimates of the $b$-vector can have a substantial error as the number of noise samples increases. 
This, in turn, results in inaccurate state estimation using the decomposition.
In contrast, the error in the reconstruction of the density matrix using the bilevel approach \eqref{SDPdensity3} in the Frobenius norm (in the rightmost column) does not increase with the number of samples of noise as fast as in the decomposed approach. 
In particular, in a low-noise regime (where for each projection operator, there are up to 75 samples of noise admixed to 5,000 measurements), the bilevel approach seems very robust.

\begin{table}[h!]
\vskip 12pt
\resizebox{\columnwidth}{!}{
\begin{tabular}{ll|rrr|r}
\hline
Noise  & $b$  & $b_x$   & $b_y$   & $b_z$ & Error  \\ \hline
0  & Orig.     & -0.0008 & -0.4674 & -0.902 & 0 \\
0 & Est.    & -0.0011  & -0.466  & -0.898 & 0.0022 \\
10 & Est.   & -0.0012 & -0.4616 & -0.892 & 0.0015 \\
25 & Est.   & -0.0036 & -0.4636 & -0.8856 & 0.0051 \\
50 & Est.   & -0.0032 & -0.458 & -0.892 & 0.0026 \\
75 & Est.   & -0.004 & -0.4634 & -0.878 & 0.0088 \\
100 & Est.  & -0.9952 & -0.9864 & -0.8864 & 0.5021 \\
150 & Est.  & -0.998 & -0.992 & -0.8815 & 0.5043 \\
250 & Est.  & 0.9916  & 0.9852 & -0.8816 & 0.5026 \\
500 & Est.  & 0.9948  & -0.9892 & -0.8784 & 0.5044 \\
750 & Est.  & 0.936   & -0.992 & -0.8828 & 0.4870 \\
1000 & Est. & -0.992  & -0.9932  & -0.8888 & 0.5005 \\
1500 & Est. & 0.9948  &  0.994 & 0.8896 & 1.3222 \\ 
2000 & Est. & 0.994   & 0.99  & -0.8924 & 1.3226 \\  \hline
\end{tabular}
}
\caption{
Error in the estimation of the $b$-vector as a function of the number of samples of Gaussian noise admixed to 5,000 measurements per projection operator: 
The first row denotes the original value of the $b$ vector (Orig.) while the subsequent rows (Est.) present estimates thereof, with 
the estimates of the b-vector obtained using an EM algorithm  displayed in columns $b_x$, $b_y$, and $b_z$.
The error in the reconstruction of the density matrix using the bi-level approach \eqref{SDPdensity3} in Frobenius norm is displayed in the right-most column. 
}
\label{T1}
\end{table}

\section{Extensions}

One could apply a similar bilevel view to a number of related problems. For relevant work, in relation to polynomial optimization methods, see Ref. \cite{Bondar2022}.

\subsection{Quantum Hamiltonian Identification}

The state of a quantum system, such as the superconducting qubit in which we are interested, evolves in time from the input Hilbert space $\mathcal{H}_{\rm in}$ to the output Hilbert space $\mathcal{H}_{\rm out}$ according to a quantum Hamiltonian operator $H: \mathcal{H}_{\rm in} \to  \mathcal{H}_{\rm out}$ that satisfies the Liouville evolution equation:
\begin{equation}\label{Lou}
    \frac{d\bs \rho(t)}{dt} = - \frac{\im}{\hbar} [H, \bs \rho(t)].
\end{equation}
When considering the discrete-time evolution of $\bs \rho_{i}$ at time $t=i$ to $ \bs \rho_{i+1}$ at time $t=i+1$, the discrete analog of Eq. \eqref{Lou} can be written as follows using the Kraus map (Kraus operator sum representation):
\begin{align}
    \bs \rho_{i+1} &= \mathcal{E}(\bs \rho_i) \\
               &= \sum_{k=1}^{d^2-1} \bs E_k^{} \bs\rho_{i} \bs E_k^{\dagger},
\end{align}
where $\sum_{k=1}^{d^2-1}\bs E_k^\dagger \bs E_k = \bs 1$ and $\mathcal{E}$ denote the unknown quantum operation responsible for the evolution of the density matrix. To perform quantum Hamiltonian identification (QHI), we sample the unknown process $m$ times, resulting in output state trajectories indexed by the lower index (see below). We introduce the following notation for the output and estimated density-matrix trajectories:
\begin{equation}
\begin{aligned}
    {\bs\rho}_{\rm out}^{(j)}(t=i) &\equiv {\bs\rho}_i^{(j)} \\
    {\bs\rho}_{\rm est}^{(j)}(t=i) &\equiv \widehat {\bs\rho}_i^{(j)} \ 
\end{aligned}
\end{equation}
where trajectory $j \in \{1,\ldots, N\}$. The index $i$ here denotes the ordinal number of the sample (discrete time). We model the evolution of the states as a linear dynamical system: 
\begin{equation}
\begin{aligned}\label{LDS}
{\bs\rho}^{(j)}_{i} &= \bs G  {\bs\rho}^{(j)}_{i-1}, \\
\widehat {\bs\rho}^{(j)}_{i} &= \bs J  {\bs\rho}^{(j)}_{i}. \
\end{aligned}
\end{equation}
Here $\bs G,\bs J$ are the system matrices that we are interested in recovering. Effectively, $\bs G$ corresponds to the evolution matrix (Hamiltonian), while $\bs J$ is a matrix that transforms the hidden state to the observed state measured by the apparatus.
To this end, we define the loss function:
\begin{equation}\label{loss}
    f_{\rm loss} = \sum_{i,j} \norm{{\bs\rho}_i^{(j)} - \widehat {\bs\rho}_i^{(j)}}^2_F 
\end{equation}
Using the Kraus operator sum representation with fixed basis $\{ \bs E_k\}_{k=1}^{d^2-1}$ in the space of Hermitian matrices $\mathbb{H}_1^{N}$, a first physical formulation of the state estimation problem, in terms of an SDP, takes the following form:

\begin{equation}
\label{Denys_sdp}
\begin{aligned}
    \min_{U}    & \quad f_{\rm loss}, \\ 
    {\rm  s.t.} & \quad {\bs\rho}_{i+1} = \sum_{k=1}^{d^2-1}\bs E_k{\bs\rho}_i \bs E_k^\dagger\\
                & \quad \sum_{k=1}^{d^2-1}\bs E_k^\dagger \bs E_k=\bs 1,\
\end{aligned}
\end{equation}
where $\bs U = \{ \bs{\rho}_i^{(j)},\bs E_{k} \}$. By algebraic manipulations one can show that $\sum_{k=1}^{d^2-1} \bs E_k^*\bs E_k =\bs  G$ and, as a result, Prob. \eqref{Denys_sdp} is reformulated as:
\begin{equation}
\label{primal_QHI}
\begin{aligned}
    \min_{S}    & \quad f_{\rm loss}, \\ 
    {\rm  s.t.} & \quad \rho^{(j)}_{i}  = \bs G  \rho^{(j)}_{i-1}, \\
                & \quad b^{(j)}_{i} = \bs J  \rho^{(j)}_{i}, \\ 
                & \quad \widehat {\bs\rho}^{(j)}_{i}  = \QST(b^{(j)}_{i}), \\
                & \quad {\bs\rho}_i^{(j)}, \widehat {\bs\rho}_i^{(j)} \in \mathbb{S}_+^{2} \text{ for each } i,j \\
                & \quad \trace{ \bs\rho^{(j)}_{i}} = \trace{\widehat {\bs\rho}^{(j)}_{i}} = 1. \
\end{aligned}
\end{equation}
Note that the index $i\in \{1,\ldots,N\}$ in $b_i$ corresponds to discrete time and should not be confused with $I \in \{x,y,z\}$. Therefore, we show that QHI can also be expressed as a bilevel SDP where the lower-level problem is Problem (\ref{SDPdensity2}), the main object of study in this article. 

\section{Conclusions}
 
We have considered, for the first time, a joint problem of quantum state tomography and discriminating between the states of a quantum system using the signal actually obtained in the dispersive readout, or similar mechanisms.
This allows for lower sample complexity of quantum state tomography compared to traditional approaches, which discriminate first and perform state estimation second, while achieving the same error in the estimate of the state. 
Considering robust statistics \cite{huber2004robust}
in this context allows many important extensions. 
 

\begin{acknowledgments}
We wish to acknowledge Denys Bondar, Zakhar Popovych as well as Christos Aravanis for helpful discussions.
Our work has been supported by OP VVV project CZ.02.1.01/0.0/0.0/16\_019/0000765 ``Research Center for Informatics''.
\end{acknowledgments}

\bibliographystyle{unsrt}
\bibliography{main}

\clearpage
\appendix

For the convenience of the reader, we provide additional background material on mathematical optimization and the dispersive readout, 
as well as further numerical illustrations in the supplementary material.

\section{Further background on mathematical optimization}

\subsection{Semidefinite Programming}
Semidefinite programming corresponds to optimization problems where the objective function is a linear function that involves a positive semidefinite matrix and constraints are given as an intersection of the convex cone of positive semidefinite matrices and an affine subspace \cite{d254b5a55d4841d68f4c435f6170c1b2}. Let us recall the most basic definitions:
Consider a matrix $X \in \mathbb{S}^n$, that is a $n\times n$ symmetric matrix and let $C(X)$ a linear function of $X$:
\begin{align}
    C(X)  & = \braket{C,X} \\
          & = \trace(C^\top X) \\
          & = \sum_{i=1}^n\sum_{j=1}^n C_{ij}X_{ij}.
\end{align}

\begin{definition}
    A primal SDP is a convex optimization problem with data consisting of a symmetric matrix $C$ and $m$ symmetric matrices $A_1,\ldots,A_m$, as  well as the $m$-dimensional vector $b$. One looks for a feasible solution $X$. The optimum is denoted as $p^*$. The problem takes the form:
\begin{equation}\label{pSDP}
    \begin{aligned}
         \textup{minimize}  & \qquad \braket{C,X} \\
         \textup{such that} & \qquad \braket{A_i,X}=b_i, \\
                         & \qquad  \, X\succeq 0.
    \end{aligned}
\end{equation}
\end{definition}

\begin{definition}
    The dual SDP to the primal SDP (\ref{pSDP}) is a convex optimization problem with data consisting of the same symmetric matrix $C$, same $m$ symmetric matrices $A_1,\ldots,A_m$, same $m$-dimensional vector $b$ as well as an $m$-dimensional vector vector $y$ and a matrix $S$. One seeks a feasible solution $(y,S)$ and the optimum is denoted as $d^*$. The problem takes the form:
\begin{equation}\label{dSDP}
    \begin{aligned}
        \textup{minimize}  & \qquad \sum_{i=1}^m y_ib_i  \\
        \textup{such that} & \qquad \sum_{i=1}^m y_iA_i + S = C, \\
                         & \qquad  \, S\succeq 0.
    \end{aligned}
\end{equation}
\end{definition}

\begin{definition}
If feasible solutions $X$ for the primal SDP and $(y,S)$ for the dual SDP exist, then the duality gap is defined as: 
\begin{align}
    \braket{C,X} - \sum_{i=1}^m y_ib_i \geq 0.
\end{align}
\end{definition}

\subsection{The standard SDP formulation of QST}

The shape-constrained least-squares (introduced in the main article)
\begin{align}\label{IBMopt}
 \norm{\bs A\vect(\widehat {\bs \rho}) - b}_{2}^2,
\end{align}
can be solved using first-order algorithms. 
Nevertheless, in this article, we will reformulate it as a semidefinite programming (SDP) problem \cite{wolkowicz2012handbook}:

\begin{equation}\label{SDPdensityAPP}
\begin{aligned}
    \QST_{\bs A}(b) := \arg \min_{\quad {\widehat{\bs\rho}}\, \succeq 0} &\quad \norm{\bs A \vect(\widehat{\bs\rho}) - b}_{2}^2 \\
    {\text{subject to}} & \quad \trace{(\widehat{\bs\rho})} = 1 \\
                        & \quad \widehat{\bs\rho} \in \mathbb{S}_+^2, \
\end{aligned}
\end{equation}
where $\mathbb{S}^n_+$ denotes the space of  complex-valued positive-semidefinite symmetric matrices. Note this is precicely the same as Eq. \eqref{SDPdensity}.
Solving the SDP \eqref{SDPdensityAPP} can be seen as a map from the space of recorded measurements $b$ to the space of estimates $\widehat{\bs \rho}$,
as suggested in Equation (4) in the main body.
Prob. (\ref{SDPdensityAPP}) easily generalizes to (i) other measurement-basis choices as well as to (ii) higher-level systems by the corresponding generalization of $\bs A$ and $b$.
The convexity is preserved, as the subspace of all density matrices embedded in the space of all Hermitian operators that act on a Hilbert space $\mathcal{H}$ forms a convex subspace. 

\subsubsection*{Evaluation map}
In Prob. (\ref{SDPdensity}), we show how to construct the matrix $\bs A$ for any measurement basis, but we have to assume that $b$ is given. Generically, the $b$ vector can be seen as a function from the space of the possible quantum-device measurements $\Omega_{\bs M}$ for the observable $\bs M$ to the interval of empirical estimates $B:=[-1,1]$. Let us denote by $\Omega_{\bs M}^B$ the set of all maps $f_j : \Omega_{\bs M}\to B$, parametrized by $j\in \mathbb{Z}$. Then $b_I := {\rm ev}(f_j,\{s_1,\ldots, s_{n_I})$, where $\{s_1,\ldots, s_{n_I}\}
\in \Omega_{\bs M}$ and ${\rm ev}$ is the evaluation map. In the following subsection, we will pick an element of $\Omega_{\bs M}^B$, for $\bs M \in \{\sig_x,\sig_y,\sig_z \}$, that allows us to re-formulate Problem (\ref{SDPdensity}) as a bilevel problem, i.e., an optimization problem that involves at least one constraint to the optimizers of another well-defined optimization problem.

\subsection{Bi-level optimization}
Following \cite{dempe_dutta_2010} a bilevel optimization problem can be considered as the following optimization problem:

\begin{equation}\label{A}
    \begin{aligned}
        \text{minimize} & \qquad F(x,y) \\
        \textup{such that} & \qquad G(x) \leq 0 \\
                          & \qquad  y \in \arg \min_y   f(x,y) \\
                          & \qquad  \text{subject to } g(x,y) \succeq 0
    \end{aligned}
\end{equation}

\noindent
Here, $F,f:\mathbb{R}^n\times \mathbb{R}^m \to \mathbb{R}$, also $G: \mathbb{R}^n \to \mathbb{R}^k$ and $g \in \mathbb{S}_+^p$, the space of $p\times p$ PSD matrices. Problem \eqref{A} is an optimization problem called the upper-level whose constraint region is determined implicitly by the graph of the solution set mapping of another mathematical optimization problem, the lower-level problem which is defined as
\begin{equation}\label{B}
    \begin{aligned}
        \text{minimize} & \qquad f(x,y) \\
        \text{subject to} & \qquad  \text{subject to } g(x,y) \succeq 0
    \end{aligned}
\end{equation}

For $x\in \mathbb{R}^m$, $y \in \mathbb{R}^n$ and $\Omega \in \mathbb{S}^p$, the space of $p\times p$ symmetric matrices, one can introduce the Lagrangian:
\begin{align}
    L(x,y,\Omega) := f(x,y) + \braket{\Omega, g(x,y)}
\end{align}
and recall that the lower-level feasible set is $Y = \{ y \in \mathbb{R}^n | g(x,y) \in \mathbb{S}_+^p \}$ and solution set mapping $\Psi(x) = \arg \min_y \{ f(x,y) | y\in Y(x)\}$. The set of regular Lagrange multiplier matrices of (\ref{A}) is defined as:
\begin{equation}
    \begin{aligned}
    \Lambda(x,y) &:= \{\Omega \in \mathbb{S}^p | 0=\nabla_y L=\braket{\Omega,g} \\
    &\quad \text{ such that } g\in\mathbb{S}_{+}^p,  \Omega \in \mathbb{S}_{-}^p \}.
\end{aligned}
\end{equation}

\begin{assumption}
    The functions $F$ and $G$ are continuously differentiable, while $f$ and $g$ are twice continuously differentiable. For any $x\in Q$, where $Q = \{ x \in \mathbb{R}^n | G(x,y) \leq 0\} $, the map $y \mapsto f(x,y)$ is convex, whereas the map $y \mapsto g(x,y)$ is $\mathbb{S}_{-}^p$-convex. Moreover, we assume 
    \begin{align}
        0&= \nabla_y \braket{\Omega,g(x,y)}, \\
        0&= \braket{\Omega, g(x,y)}, \\
        \Omega & \in \mathbb{S}_{-}^p \Longleftrightarrow \Omega =  0. \\
    \end{align}
\end{assumption}
Using \cite[Prop. 3.2]{bonnans:inria-00073819}, we see that
\begin{align}
    \forall (x,y) \in Q \times \mathbb{R}^m, \,\,\, y\in \Psi \Longleftrightarrow \Lambda(x,y) \neq \emptyset. 
\end{align}
Here, $\Psi : \mathbb{R}^n \to \mathbb{R}^m$ is the solution map of the lower-level problem:
\begin{align}
    \Psi := \{ f(x,y) | g(x,y) \in \mathbb{S}_+^p \}.
\end{align}
Moreover, for any $(x,y) \in {\rm gph}\Psi \cap (Q \times \mathbb{R}^m)$ the set $\Lambda(x,y)$ is non-empty, convex, and compact. This justifies the substitution of the lower-level problem with the KKT conditions into Prob. \eqref{A}:

\begin{equation}
    \begin{aligned}
        \min_{x,y,\Omega} & \qquad F(x,y) \\
        \text{such that} & \qquad G(x,y) \leq 0 \\
                         & \qquad \nabla_y L = 0 \\
                         & \qquad g(x,y) \succeq 0 \\
                         & \qquad \Omega \in \mathbb{S}_{-}^p \\
                         & \qquad \braket{\Omega, g(x,y)} =0. \\
    \end{aligned}
\end{equation}
In \cite{dempe_dutta_2010}, the authors show that a classical bi-level programming problem and its KKT
reformulation are equivalent with respect to global optimal solutions, whereas these problems
do not need to coincide with respect to local optima.

\section{Further background on the dispersive readout}\label{app:readout}

\begin{figure}[h!]
    \centering
    \includegraphics[width=1.0\columnwidth]{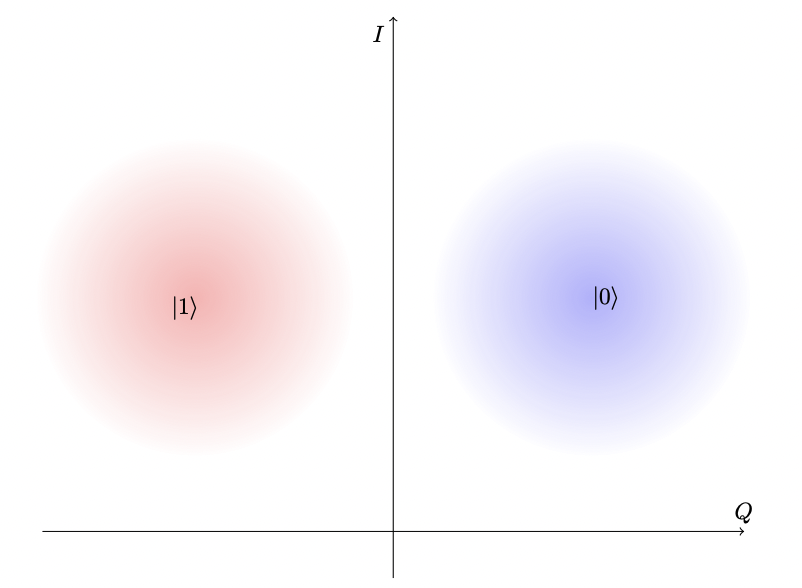}
    \caption{The I-Q plane data corresponds the phase space measurements of the response signal.
    While this figure is idealized, in reality, the scatter plot contains a lot of noise and robust techniques are required. Note that we can essentially identify the pointer basis $\{ \ket{g}, \ket{e}\}$ to the standard qubit state basis $\{\ket{0}, \ket{1} \}$. The measurement described above has been performed in a fixed projective observable, say $\sig_z$. The process needs to be repeated for the other two observables, $\sig_x, \sig_y$ in order to get a full description of the state.}
    \label{IQ}
\end{figure}

In this section, we provide further details on the dispersive readout and the IQ-plane. We want to remind the reader that the notion of discrimination refers to the process of determining whether the qubit was measured in the $\ket{0}$ or the $\ket{1}$ eigenstate with respect to the measurement operator $\bs M$ as discussed in Sec. \ref{sec1:level1}. In the dispersive readout, the readout chain is composed of the three levels of output data: level 0 \emph{raw data}, level 1  \emph{I-Q plane data} and level 2 \emph{discriminated data}.
The output response pulse (at level 0) can be mapped to a complex number (at level 1) that can be decomposed as the amplitude response $I$ and the phase response $Q$ and its precise meaning is explained below. Repetition of the measurement $n$ times yields a mixture of two distributions, as shown in Figure \ref{IQ}. 

As described in Sec. \ref{sec:sec2}, the measurement process in a device such as the transmon qubit consists of probing the resonator with a pulse of frequency $\omega_{\text{probe}}$. The maximum fidelity is achieved when $\omega_{\text{probe}} = (\omega_C + \omega_{C}^{\chi})/2$. A short readout pulse $s_{\text{r.o.}}(t)$ is then directed towards the resonator to interact with it, and thus interact with the qubit and be transmitted back to the control line. Assuming a linear pulse, its readout waveform reads:
\begin{align}
    s_{\rm r.o.}(t) = A_{\rm r.o} \Big(\cos (\omega_{\rm r.o}t) + \vartheta_{\rm r.o.}\Big),
\end{align}
where $A_{\rm r.o}$ is the amplitude of the probe pulse and $\theta_{\rm probe}$ is the phase, both of which depend on the state of the qubit. We can rewrite the waveform as
\begin{align}
    s_{\rm r.o.}(t) &= {\rm Re}\left(A_{}  e^{i(\omega_{\rm r.o.}t + \vartheta_{})} \right)  \\
         &= {\rm Re} \left(A_{} e^{\im \theta_{}} e^{\im \omega_{\rm r.o.}t} \right)
\end{align}
where we skipped the labels on the amplitude, frequency, and phase of the readout pulse. The quantity $s|_{\omega_{\rm r.o.}} = A_{} e^{\im \theta_{}}$ is called a phasor, and for a fixed frequency it completely specifies the pulse. The qubit resonance readout is performed by recording the \emph{in-phase} component $I$ and the \emph{quadrature} component $Q$ of the phasor:
\begin{align}\label{IQeq2}
    s|_{\omega_{\rm r.o.}} &= A_{}\Big( \cos(\theta_{}) + \im \sin(\theta_{})\Big) \\
                            &= I + \im Q, \
\end{align}
i.e., Eq. \eqref{IQeq}.
The I-Q plane $\simeq \mathbb{C}$ can be thought of as the phase space of the resonator-qubit coupled system. Once the signal has been transmitted back from the resonator to the control line, a comparison of the readout pulse to the original pulse is performed. Using the phase-shift between the known input pulse and the measured output, the qubit state can be mapped onto the complex I-Q plane.

Thus, a single pulse sent to the qubit maps to a point in the the complex I-Q  plane. In total, $N$ repetitions of the measurement provide a distribution that allows one to create a histogram of the recorded events and assign the probabilities of experiment outcomes. Repetition of the same procedure, with further copies of the state measured with different measurement operators of the $n$-dimensional measurement basis (albeit possibly with different pulse frequencies) allows one to study a $n$-dimensional histogram from which one can use to proceed to the estimation of the measured state.

\begin{figure}[h!]
    \centering
    \hspace*{-0.4cm}
    \includegraphics[width=1.0\columnwidth]{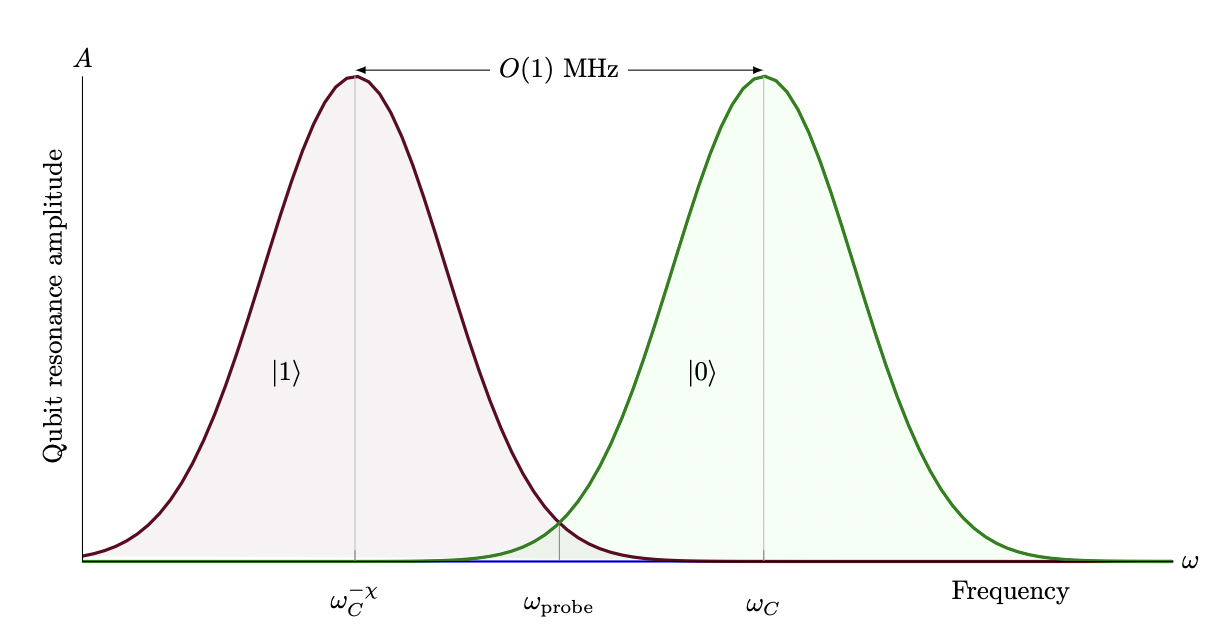}
    \caption{Resonance frequency-amplitude diagram for the qubit. The input probe pulse $\omega_{\rm probe}$ excites the readout to either the green Lorentzian peak corresponding to $\ket{0}$ or the blue Lorentzian peak corresponding to $\ket{1}$. The sign of the phase difference of $\omega_{\rm probe}$ to $\omega_{\rm r.o.}$, the readout frequency, will reveal the pointer state the qubit is (with respect to one of the Pauli observables). The difference between the two resonance peaks is approximately $2\chi$ and it is of the order of a few MHz.}
    \label{energy}
\end{figure}

\section{Details of our bilevel formulation}
\label{sec:bilevelSDP}

\subsection{Known parameters and positive adversarial noise}

When the mixture model is contaminated by noise, that is, $\alpha_2 > 0$, we need to consider a constrained optimization problem. This, in general, does not have a unique solution, and we need to consider the inclusion
$b_I \in [\IQP_{\mathcal{D}_{\ket{0},\ket{1}}}(x)]_I.$
While in theory, one could consider the nonconvex non-commutative polynomial optimization problem (10) from the main article, and consider the first-order optimality conditions of their (globally convergent) SDP relaxations \cite{Navascu_s_2008,nie2020lagrange}, a simpler approach is to consider a continuous relaxation in variables $U = \{f_s^{\ket{0}}, f_s^{\ket{1}},f_s^{\rm noise} \}\in [ 0, 1 ]$ over the polyhedron defined by Equation (10) from the main article.
This convex optimization problem can be replaced by first-order optimality conditions under mild assumptions \cite{nie2020lagrange}.

\subsection{Unknown means and known covariance matrices}

Let us recall Equation (9) from the main article:

\begin{align}
\label{dis_adv}
 f(u_s) & = \sum_{i=\{1,2\}}   \mathcal{D}_{\ket{i}}(u_s)  +\alpha_2 g(u_s) \\
\textrm{where } \mathcal{D}_{\ket{i}}(u_s) & :=  \frac{1}{\sqrt{2\pi}} \frac{\alpha_i}{|\Sig_i|^\frac{1}{2}} e^{(u_s - {\mu}_i)^\top \Sig_i^{-1}(u_s-{\mu}_i)}, \notag
\end{align}

Assume that the mean vectors $\mu_i$ of Eq. \eqref{dis_adv} are unknown, but the covariance matrices ${\Sig}_i$ are known. 
In this case, one has to estimate the mean vectors $\hat{\mu}_i$ as well as mixture weights $\hat{a}_i$ the lower-level problem (10) from the main article corresponds to a nonconvex, but commutative POP. Although this makes the problem nontrivial, it has been studied \cite{jeyakumar2016convergent}. Furthermore, the parameter space can be reduced by making the reasonable assumption $\mu_0^1 \approx - \mu_1^1, \mu_0^2 \approx  \mu_1^2$, for $\mu_i = (\mu_i^1\,\,\, \mu_i^2)^\top$. 
In particular, \cite[Theorem 4.7]{jeyakumar2016convergent} shows that assuming the Mangasarian-Fromovitz constraint qualification (or, less strictly, 
that there exists a representation of the feasible set of the lower-level problem as a finite union of closed convex sets with nonempty interiors), there exists an $\epsilon_0$, such that for all 
$\epsilon \in [0, \epsilon_0)$, 
one can obtain an $\epsilon$-approximation of the problem by a convexification, which turns out to be an SDP that could be utilized, considering the recent study \cite{doi:10.1137/16M1099303} of bilevel optimization with an SDP at the lower level.

The dimension of this SDP will grow rapidly with $\epsilon$, but this is justified by the well-known issues \cite{NIPS2016_3875115b} in estimating the parameters of a Gaussian mixture model using the EM algorithm, which would be the straightforward alternative. 
As a practically relevant alternative, one may consider the first available SDP within the hierarchy, which resembles Shor's \cite{shor1987quadratic,shor1990dual} SDP relaxation and its KKT conditions. 

\subsection{Unknown means and unknown covariance matrices}

Finally, let us assume that both the mean vectors and the covariance matrices of Equation (\ref{dis_adv}) are unknown and there is contamination ($\alpha_2 > 0$). The lower-level problem (10) from the main article then corresponds to a nonconvex NCPOP. One can solve such a bilevel polynomial optimization problem with a nonconvex lower-level problem using hierarchies of semidefinite programming relaxations. %
Hierarchies of SDP relaxations of the NCPOP,  such as the NPA hierarchy \cite{Navascu_s_2008}, essentially convert the original NCPOP to a series of SDP problems labeled by $k$ such that for some $k$, the optimum of the SDP converges to the optimum of the NCPOP.  Once the NCPOP is convexified in the form of an SDP, one can employ the KKT conditions in a manner similar to \cite{jeyakumar2016convergent}.\\

\section{Further numerical illustrations}
In this section, we provide a numerical illustration of the method proposed in the main article. 

Consider a two-level system in state:
\begin{align} \label{truestate}
     \bs \rho_{22} &= \begin{bmatrix}
        0.056  &   \im 0.229 \\
        \im0.229  & 0.944 \ \end{bmatrix}
\end{align}
where we would like to estimate the eigenstate counts for the Pauli observables.
Given a Gaussian mixture model for each $I\in \{x,y,z\}$ with mean vectors $\mu_0, \mu_1$ and covariance matrices $\bs \Sigma_0, \bs \Sigma_1$, we can sample data that resemble the IQ-plane data using the state counts:
\begin{align}\label{fake_data} \nonumber
    n_x^{\ket{0}} &= 4996, \qquad n_x^{\ket{1}} = 5004 \\ 
    n_y^{\ket{0}} &= 2663, \qquad n_y^{\ket{1}} = 7337 \\ \nonumber
    n_z^{\ket{0}} &= 540,\,\,\,  \qquad n_z^{\ket{1}} = 9460. \
\end{align}
We consider two spherical Gaussians of means $\mu_0 = \begin{pmatrix} 2.5 & 2.0 \end{pmatrix}^\top$,\,\, $\mu_1 = \begin{pmatrix} -2.5 & 2.0 \end{pmatrix}^\top$.
We sample the first Gaussian with frequency $n_I^{\ket 0}/n_I$ and the second Gaussian with frequency $n_I^{\ket{1}}/n_I$, $I \in \{x,y,z\}$. We perform $n=10,000$ measurements of the state with respect to each of the elements in the Pauli basis to obtain values for the empirical estimates of the expectation values of the Pauli observables:
\begin{align} \label{Qutipb}
    b = \begin{bmatrix} -0.0006 & -0.4674 & -0.892 \end{bmatrix}^\top.
\end{align}


\begin{center}
\renewcommand{\arraystretch}{0.6}
\begin{table}[b!]
\resizebox{\columnwidth}{!}{
\begin{tabular}{ c | c| c} \hline \hline
   & $\hat \mu_0$ & $\hat \mu_1$ \\  \hline
 $\sigma_x$ & $\SmallMatrix{2.49752013 &  1.98083953}^\top$  &$\SmallMatrix{-2.50895142 &  1.96288668}^\top$  \\  
 $\sigma_y$ & $\SmallMatrix{2.48478612 &  1.99784236}^\top$ & $\SmallMatrix{-2.55945933 & 1.96636545}^\top$ \\
 $\sigma_z$&$\SmallMatrix{2.49009736 &  1.99746892}^\top$& $\SmallMatrix{-2.60719554 &  1.92029166 }$  \\
 \hline  
\end{tabular}}
\caption{Mean vector estimates for each of the three spherical Gaussian mixture models using the EM algorithm.}
\label{tab:s1}
\end{table}
\end{center}

\begin{center}

\begin{table}[b!]
\resizebox{\columnwidth}{!}{
\begin{tabular}{ c | c| c} \hline \hline
   & $\widehat{\Sig}_0$ & $\widehat{\Sig}_1$ \\  \hline
 $\sigma_x$ & $
\SmallMatrix{0.9943854 & -0.0298129 \\ -0.0298129 & 0.987068 }
$ & $
\SmallMatrix{ 0.9856472 &  0.009437 \\  0.009437 &  0.9744658}
$ \\  
 $\sigma_y$ & $\SmallMatrix{1.0076411 & -0.0044255 \\ -0.0044255 & 0.994228}$ &$\SmallMatrix{0.9347116 &  0.0104711\\ 0.0104711 &  0.9281853}$ \\
 $\sigma_z$&$\SmallMatrix{1.0082151 &  0.0014959 \\ 0.0014959 &  0.9844126
}$ & $\SmallMatrix{ 1.0102680 & -0.1381553 \\ -0.1381553 & 0.9478898}$  \\
 \hline  
\end{tabular} }
\caption{Covariance matrix estimates for each of the three spherical Gaussian mixture models using the EM algorithm.}
\label{tab:s2}
\end{table}
\end{center}

\begin{table}[b!]
    \centering
    \begin{tabular}{c|| c|c|c}\hline \hline
                     & $b_x$ &$b_y$ & $b_z$ \\ \hline
        \texttt{Qutip} &-0.0006 & -0.4674 & -0.8920 \\  \hline
        \text{ME} &-0.0044 & -0.4659 & -0.8979  \\ \hline
        \text{SDP}&-0.0004  &   -0.4480 &  -0.8913 \\ \hline
    \end{tabular} 
    \caption{Estimates for the $b$ vector.}
    \label{tab:s3}
\end{table}

Traditionally, having the IQ-plane data, one would use EM algorithm for half of the dataset for calibration purposes, and then use the calibration results to decide the membership of the rest of the measured points. Using the counts \eqref{fake_data} it is trivial to assign the estimates of the expectations of the measurement operators and to obtain the $b$ vector:
\begin{align}
    b = \begin{bmatrix} -0.0044 & -0.4659 & -0.8979 \end{bmatrix}^\top.
\end{align}
which approximates the $b$-vector that one obtains from \texttt{Qutip} \eqref{Qutipb}. Subsequently, the optimizer of the standard SDP formulation of QST, Equation (7) from the main article, is:
\begin{align}
    \bs \rho_{(12)} = \begin{bmatrix}
        0.0597     &  -0.0008 + 0.2274\im \\
         -0.0008 - 0.2274\im  & 0.9403    \
    \end{bmatrix},
\end{align}
which, given the small supply of samples to the algorithm, is a good estimate of the true state \eqref{truestate}. For comparison, using \eqref{Qutipb} we obtain a state estimate:
\begin{align}
    \bs{\rho}_{\texttt{Qutip}} =  \begin{bmatrix}
        0.0571  & -0.0003 +  0.2321\im \\
        -0.0003 - 0.2321\im  & 0.9429 \
    \end{bmatrix}.
\end{align}
Using our approach, Equation (8) from the main article, which utilizes the IQ-plane data directly,  we explicitly count states using Equation (10) from the main article,  to estimate the true state $\bs \rho$  \eqref{truestate} as:
\begin{align}
     \bs{\rho}_{(24)}  = \begin{bmatrix}
       0.0544     &    -0.0002 + 0.2240\im \\
          -0.0002 - 0.2240\im  & 0.9456  \
    \end{bmatrix}.
 \end{align}   

The Frobenius norm of the difference, as computed by MATLAB\textsuperscript \textregistered, is $
\|\bs{\rho}_{\texttt{Qutip}}- \bs{\rho}_{(12)}\|_F = 0.6455$ and 
$\|\bs{\rho}_{\texttt{Qutip}}-\bs{\rho}_{(24)}\|_F =  0.6406$, respectively, 
suggesting a modest improvement in the estimate of the quantum state. 
This numerical illustration provides only an  anecdotal evidence of the improvement that can be obtained by considering the bilevel problem. We envision that further work could corroborate the observation on real data. 

Both the standard approach to quantum state tomograpy as well as our approach, Equations \eqref{SDPdensity} and \eqref{SDPdensity2} from the main body, can be written in MATLAB\textsuperscript \textregistered using the CVX convex optimization library \cite{cvx} in the SDP mode where the algorithm, in both cases, runs very fast. The state estimation $\bs X_{12}$ requires knowledge of the $b$ vector. This is achieved with the EM algorithm (see \cite{Gupta_theoryand} for example) used on the training data obtained by sampling the Gaussians. On the other hand, the SPD that estimates $\bs X_{24}$, computes the $b$ vector as part of the problem. 



In Tables \ref{tab:s1}--\ref{tab:s3}, we present the ME-estimations for the mean vectors and the covariance matrices for the three sets of Gaussian mixtures, as well as the values of the $b$ vectors for the three different approaches.

\end{document}